\documentclass{iaus}
\usepackage{graphicx}
\usepackage[]{natbib}

\def\aj{AJ}
\def\aap{A\&A}
\def\apj{ApJ}
\def\apjl{ApJL}
\def\apjs{ApJS}
\def\mnras{MNRAS}
\def\msun{M_\odot}

\title[Modelling individual globular clusters] 
{Modelling individual globular clusters}

\author[D.C. Heggie and M. Giersz]   
{Douglas C. Heggie$^1$ and Mirek Giersz$^2$%
  }

\affiliation{$^1$School of Mathematics and Maxwell Institute
    for Mathematical Sciences, University of Edinburgh, Edinburgh, EH9
    3JZ, UK\break email: d.c.heggie@ed.ac.uk\\$^2$Nicolaus Copernicus Astronomical Center, Polish Academy of
Sciences, Warsaw, Poland
\break email: mig@camk.edu.pl}

\pubyear{2008}
\volume{246}  
\pagerange{119--126}
\date{?? and in revised form ??}
\jname{Dynamical Evolution of Dense Stellar Systems}
\editors{E. Vesperini, M. Giersz \& A. Sills, eds.}

\begin{document}

\maketitle

\begin{abstract}
Astronomers have constructed models of globular clusters for over 100
years.  These models mainly fall into two categories: (i) static
models, such as King's model and its variants, and (ii) evolutionary
models.  Most attention has been given to static models, which are
used to estimate mass-to-light ratios and mass segregation, and to
combine data from proper motions and radial velocities.  Evolutionary
models have been developed for a few objects using the gaseous model,
the Fokker-Planck model, Monte Carlo models and $N$-body models.  These
models have had a significant role in the search for massive black
holes in globular clusters, for example.  

In this presentation the problems associated with these various
techniques will be summarised, and then we shall describe new work
with Giersz's Monte Carlo code, which has been enhanced recently to
include the stellar evolution of single and binary stars.  We describe
in particular recent attempts to model the nearby globular cluster M4,
including predictions on the spatial distribution of binary stars and
their semi-major axis distribution, to illustrate the effects of about 12 Gyr
of dynamical evolution.  We also discuss work on an approximate way of
predicting the ``initial" conditions for such modelling.
\keywords{methods: numerical, globular clusters: general, globular clusters: individual (M4)}
\end{abstract}

\firstsection 

\bigskip


\section{Introduction}

\subsection{Some astrophysical questions}
There are many reasons for constructing a dynamical model of a
globular cluster, but they fall into two broad categories.  First
there are problems that can be tackled by constructing static
(equilibrium) models, such as
\begin{enumerate}
  \item Inferring the mass from the surface brightness profile, radial
		velocities,	proper motions and  mass functions.  In
		this way one can estimate the total mass of stars below the
		observational limit, such as faint white dwarfs \citep[e.g.][]{Dr1988}
\item Inferring the global mass function from local mass functions \citep[e.g.][]{Ri2004}:
   	then one can address the question of whether 
   	this is the same for all clusters.
\item Measuring cluster distances by comparison of radial velocities and 
   	proper motions:  the model is used to correct for rotation, or
   	to
   	link the different locations and stellar components which are
   	observed by the different techniques \citep[e.g.][]{Ve2006}.
\end{enumerate}

Then there is a second range of questions which require dynamic {\sl
  evolutionary} models, i.e. questions such as
\begin{enumerate}
\item Inferring the primordial mass function from local, present-day mass
   	functions: the model is used to correct for preferential
   	escape of low-mass stars \citep[e.g.][]{BM2003}.
\item Inferring primordial parameters of the binaries from their present-day 
   	statistical properties: primordial abundance, period 
   	distribution, etc \citep[e.g.][]{Kr2001}
\item  Determining the effect of dynamics on the estimation of mass
  through the virial theorem, which is affected by mass segregation
\citep{Fl2006}
\end{enumerate}
Note that these last few references do not deal with specific objects
(which is the focus of the rest of this review) but with general trends.

This paper begins with a review of the methods and observational
constraints which have been used to
construct models of individual globular clusters, often with a view to
answering the above types of question.  Then we focus on one
particular method, the Monte Carlo method, and its application to the
nearby globular cluster M4.

\subsection{Methods and constraints}

\subsubsection{The methods}
A number of techniques have been used to construct static models, to
answer questions of the first type:
\begin{enumerate}
\item Plummer's model \citep{Pl1911}
\item King's model \citep{Ki1966,PK1975}
\item Anisotropic models \citep{MB1963}
\item Multi-mass models \citep{GG1979, Pr1986, Du1990}
\item Non-parametric models \citep{GF1995}
\item Schwarzschild's method \citep{Ve2006}
\item Jeans' equations \citep{Le1992}
\end{enumerate}
Even this list may not be exhaustive.

 Much less work has been done on dynamical
evolutionary models of {\sl specific} globular clusters, but the methods
include
\begin{enumerate}
\item Gas/fluid models (\citealt{An1980} [M3])
\item Fokker-Planck models (Cohn and co-workers: \citealt{Gr1992}
 [N6624], \citealt{Du1997}
 [M15]; \citealt{Dr1993, Dr1995} [N6397]; 
	Phinney 1993 [M15])
\item Monte Carlo models (Giersz \& Heggie 2003 [$\omega$ Cen])
\item $N$-body models 
\end{enumerate}

The last of these should really not be on this list.  Though it should
be the method of choice, it has not been used for the purposes which
are the focus of this paper.  An example is the modelling of M15 by
\citet{Ba2003}, who constructed a small version which, when
scaled up, corresponded approximately to the conditions expected for
M15.  The difficulty is that unscaled $N$-body models are practically
limited to $N$ of order $10^5$ at present \citep[e.g.][]{BM2003,  Hu2007}, whereas the median for the globular clusters at the
present time is about $5\times 10^5$.  Though it might be hoped that
the gap would be bridged by the next generation of computers, it must
be recognised that all globular clusters at the present day have lost
substantial numbers of stars, and the primordial median must have been
higher.  Note that it has become possible only relatively recently to
carry out full simulations of open star clusters.  The initial mass of
M67, for instance, is estimated at about 19 000 $\msun$, i.e. about 10
times its present mass, and this simulation, with a realistic
complement of primordial binaries, took of order 1 month \citep{Hu2005}.

\subsubsection{The observational constraints}

Whichever method is chosen to model a globular cluster, there are a
number of observational constraints to be satisfied.  In historical
order of first use we have
\begin{enumerate}
\item Surface brightness and/or star counts, starting with \citet{Ze1908};
\item Radial velocities, whether central averaged values \citep{Il1976} or radial velocities of individual stars \citep{GG1979}
\item Pulsar accelerations \citep{Ph1993, Gr1993}
\item {\sl Deep} luminosity/mass functions, starting essentially with
  the advent of studies by several authors using HST (1995)
\item Accurate proper motions \citep{Bo2006}.  (We refer here to
  models built to satisfy constraints imposed by observations of
  internal proper motions, and not to the use of proper motions to establish membership.)
\end{enumerate}

\subsubsection{The Monte Carlo Model}\label{sec:mcm}

This paper focuses on an application of the Monte Carlo code developed
essentially by \citet{Gi1998, Gi2001, Gi2006}.  In this approach we assume spherical symmetry and dynamic
equilibrium, and
characterise each star by its energy $E$ and angular momentum $J$.
The code repeatedly alters 
$E$ and $J$ to mimic the effects of
gravitational encounters, 
using the theory of relaxation.  The same theory underpins
Fokker-Planck codes, and the basic Monte Carlo code provides
essentially a Monte Carlo solution of the Fokker-Planck equation.

The Monte Carlo code is rather suitable for the addition of a number of
other  process, chiefly:
\begin{enumerate}
\item The galactic tidal field, which is treated as a cutoff
\item Binaries, whose interactions are treated using cross sections
  (from \citet{Sp1987} for interactions with single stars, and
  expressions based on \citet{Mi1983,Mi1984a,Mi1984b} for interactions between binaries)
\item Stellar evolution of single stars \citep{Hu2000} and 
binary stars \citep{Hu2002}.
\end{enumerate}
Each of these requires some comment, though further details are given
by  Giersz \& Heggie in this volume, and in Sec.4 below.

\begin{enumerate}
\item There are significant differences between  a tidal field and a tidal
  cutoff, as these lead to somewhat different scalings of the
  dissolution time with $N$ \citep{Ba2001}.  We have attempted to
  mimic this with a mass-dependent lowering
of the escape energy. 
\item The Monte Carlo code described here has no triples, and so
  hierarchical triples (which are a common product of binary-binary
  encounters) have to be bypassed.  Furthermore, the use of cross
  sections hinders the inclusion of star-star collisions during 3- and
  4-body encounters.  Finally, the cross sections are not well known
  for unequal masses.
\item Stellar evolution is implemented via the McScatter interface
\citep{He2006}.  Besides the stellar evolution packages
	of Hurley et al (referenced above) it also interfaces to the
	stellar evolution package SeBa in starlab \citep{PZ1996}.  At present,
	however,   the latter is
	limited to solar metallicity, which is unsuitable for the
	study of old globular clusters.	
\end{enumerate}

In view of the above approximations and uncertainties, the testing and
calibration of the Monte Carlo code against the results of $N$-body
models (in the regime of small enough $N$) is an essential safeguard.
Such studies are described by Giersz \& Heggie in this volume.

\section{The globular cluster M4}

This, one of  the very nearest known globular clusters, was selected at the meeting
MODEST-5 (Hamilton, Canada, 2004) as a target for concerted
observational and theoretical effort, but so far little theoretical
work has been carried out.  Table 1 summarises some data for this
fascinating object.

  \begin{table}
\begin{center}
\caption{Properties of M4}
\begin{tabular}{lll}
\hline
Distance from sun	&	$^a$1.72kpc&\\
Distance from GC	&	5.9kpc&\\
Mass			&	$^a$63 000$\msun$&\\
Core radius		&	0.53 pc&\\
Half-light radius	&	2.3 pc&\\
Tidal radius		&	21 pc&\\
Half-mass relaxation time ($R_h$)	&660 Myr&\\
Binary fraction 	&	$^a$1-15\%&\\
$\left[\right.$Fe/H$]$&-1.2&\\
Age			&$^b$12Gyr&\\
$A_V$ 			&$^a$1.33&\\
\hline
\end{tabular}
\end{center}
References: All data are from the current version of the catalogue of \citet{Ha1996}, except $^a$ \citet{Ri2004} (though this is not always the original reference for the quoted
number) and $^b$ \citet{Ha2004}.
  \end{table}

The proximity of M4 makes deep observational study possible.  (See,
for example, the poster by Sommariva et al in this volume.)   For
theoretical purposes too it is well placed for study because its
binary population appears to be modest, and its initial mass may not
have been very high, as we shall see.  One complication for the Monte
Carlo code, however, is that the orbit appears to be very elliptical \citep{Di1999}, whereas we must assume a steady tidal field.

Table 2 describes the initial conditions which we adopted for this
exercise.  The primary observational data which we attempted to fit
were 
\begin{enumerate}
  \item The surface brightness profile \citep{Tr1995}
\item The radial velocity dispersion profile \citep{Pe1995}
\item The V-luminosity function (\citealt{Ri2004}, from which we
  considered the results for the innermost and outermost of their four
  annuli)
\end{enumerate}
though several other observational comparisons will be described
below.  We do not have a systematic way of arriving at a best choice
of initial parameters, though a possible approach is described towards
the end of this paper.  We began with a scaled-up version of the
models we developed for the old open cluster M67 (see Giersz \& Heggie
in
this volume), but found that a binary population of $f_b = 50$\%
tended to produce a model with too low a concentration.  Reducing the
binary concentration to 5 or 10\% produced a satisfactory surface
brightness profile, but was somewhat too massive, because of an excess
of low-mass stars, corresponding to a poor fit with the luminosity
function.  According to \citet{BM2003} it might be
possible to correct this by devising a model which lost mass at a
higher rate, but instead we elected to change the slope of the
low-mass IMF from the canonical value of $\alpha=1.3$ \citep{Kr2007} to $\alpha=0.9$.  (There is some justification for a lower value
for low-metallicity populations.)  By some
experimentation we arrived at a model which gave a fair fit to all
three kinds of observational data; see Table 3, and Figs 1-4.  Much of
the disagreement in the total luminosity is due to our assumed
distance to the cluster, which is significantly smaller than the value of 
2.2kpc given by \citet{Ha1996}, though our surface brightness profile
  is also a little faint on average.  The disagreement in the inner
  luminosity function at faint magnitudes may be attributable to the
  fact that the theoretical result assumes 100\%
  completeness, while the observational data are uncorrected  for
  completeness.  A typical plot of a completeness correction is
  given by \citet[Fig.3]{Ha2002}.

It is worth noting that no arbitrary normalisation has
been applied in these comparisons between our model and the
observations.  The surface brightness profile, for example, is
computed directly from the V-magnitudes of the stars in the Monte
Carlo simulation.

  \begin{table}
\begin{center}
\caption{Initial parameters for M4}
\begin{tabular}{ll}
\hline
Fixed parameters\\
Structure&Plummer model\\
Stellar IMF&Kroupa double power law\\
Binary mass distribution	&\citet{Kr1991}\\
Binary mass ratio               &Uniform\\
Binary semi-major axis          &Uniform in log, $2(R_1+R_2)$ to 50AU\\
Binary eccentricity             &Thermal, with eigenevolution \citep{Kr1995}\\
Metallicity $Z$&0.002\\
Age&12Gyr\\
\hline
Free parameters\\
Mass			&$M$\\
Tidal radius&$r_t$\\
Half-mass radius&$r_h$\\
Binary fraction 	&	$f_b$\\
Slope of the lower mass function&$\alpha$ (Kroupa = 1.3)\\
\hline
\end{tabular}
\end{center}
  \end{table}

  \begin{table}
\begin{center}
\caption{Monte Carlo and King models for M4}
\begin{tabular}{llll}
\hline
Quantity&MC model &MC model &King model\\
&($t = 0$)&($t=12$Gyr)& \citep{Ri2004}\\
\hline
Mass ($\msun$)&$3.40\times10^5$&$4.61\times10^4$&\\
Luminosity ($L_\odot$)&$6.1\times10^6$&$2.55\times10^4$&$6.25\times10^4$\\
Binary fraction $f_b$&0.07&0.057&0\\
Low-mass MF slope $\alpha$&0.9&0.03&0.1\\
Mass of white dwarfs ($\msun$)&0&$1.81\times10^4$&$3.25\times10^{4^\ast}$\\
Mass of neutron stars ($\msun$)&0&$3.24\times10^3$\\
Tidal radius $r_t$ (pc)&35.0&18.0&\\
Half-mass radius $r_h$ (pc)&0.58&2.89&\\
\hline
\end{tabular}
\end{center}
$\ast$: this is the quoted mass of ``degenerates''
  \end{table}


  \begin{figure}
  \begin{minipage}[]{0.45\textwidth}
{\includegraphics[height=10cm,angle=-90,width=8cm]{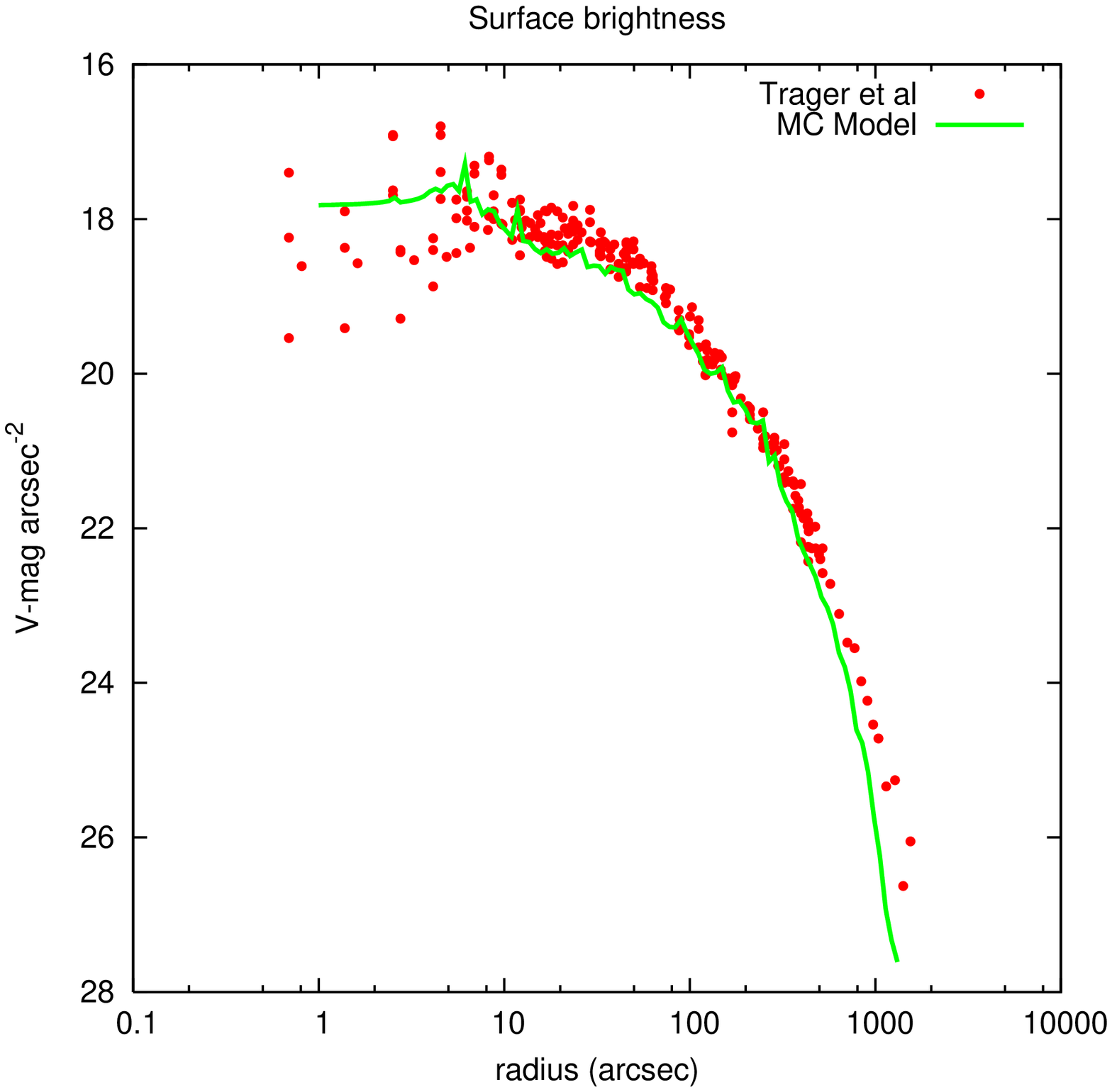}}
    \caption{Surface brightness profile of our Monte Carlo model,
    compared with the data of \citet{Tr1995}.}
\end{minipage}
\hfill
  \begin{minipage}[]{0.45\textwidth}
{\includegraphics[height=10cm,angle=-90,width=8cm]{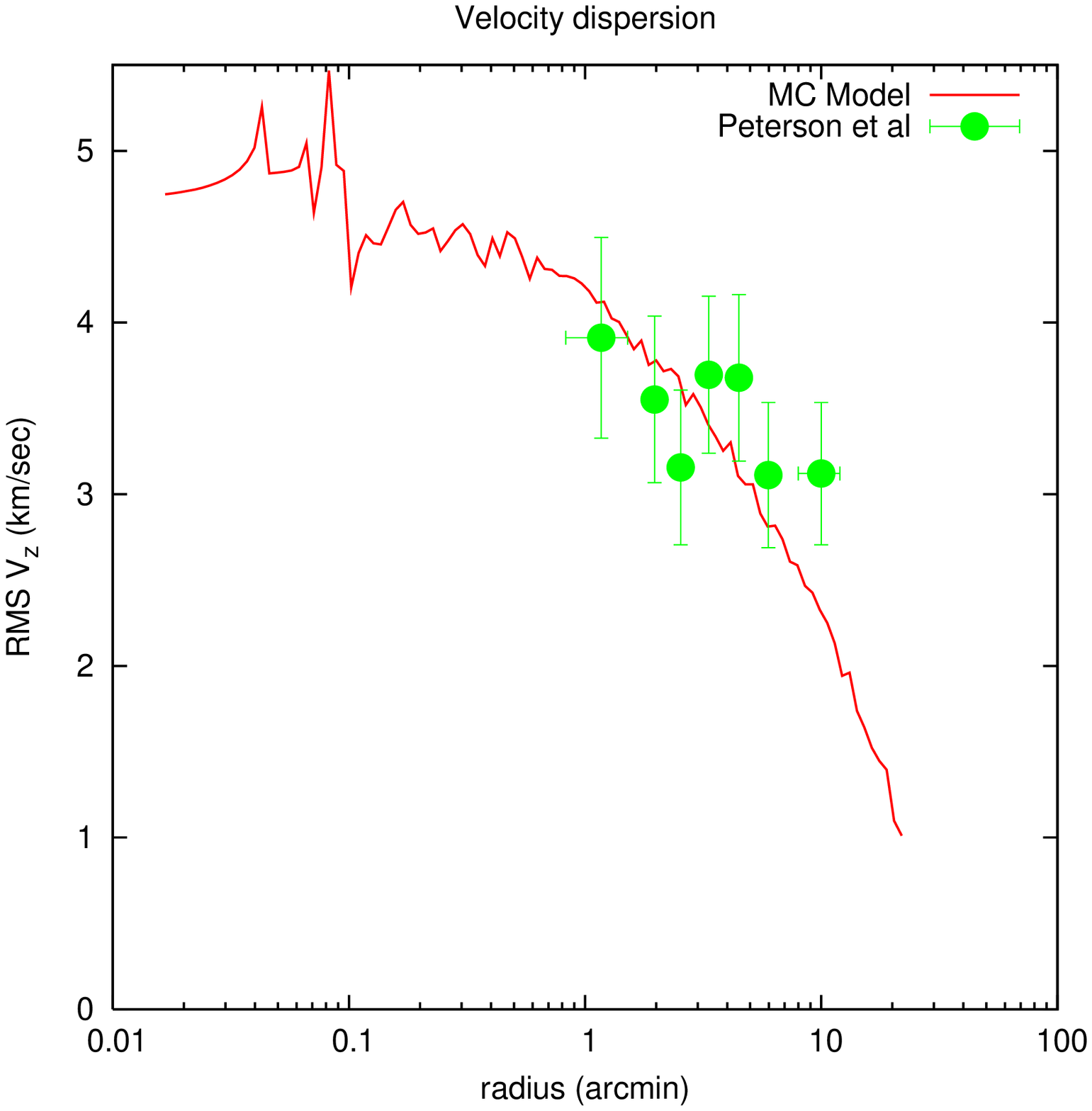}}
    \caption{Velocity dispersion  profile of our Monte Carlo model,
    compared with the data of \citet{Pe1995}.}
  \end{minipage}
\end{figure}

  \begin{figure}
    \begin{minipage}[]{0.45\textwidth}
{\includegraphics[height=10cm,angle=-90,width=8cm]{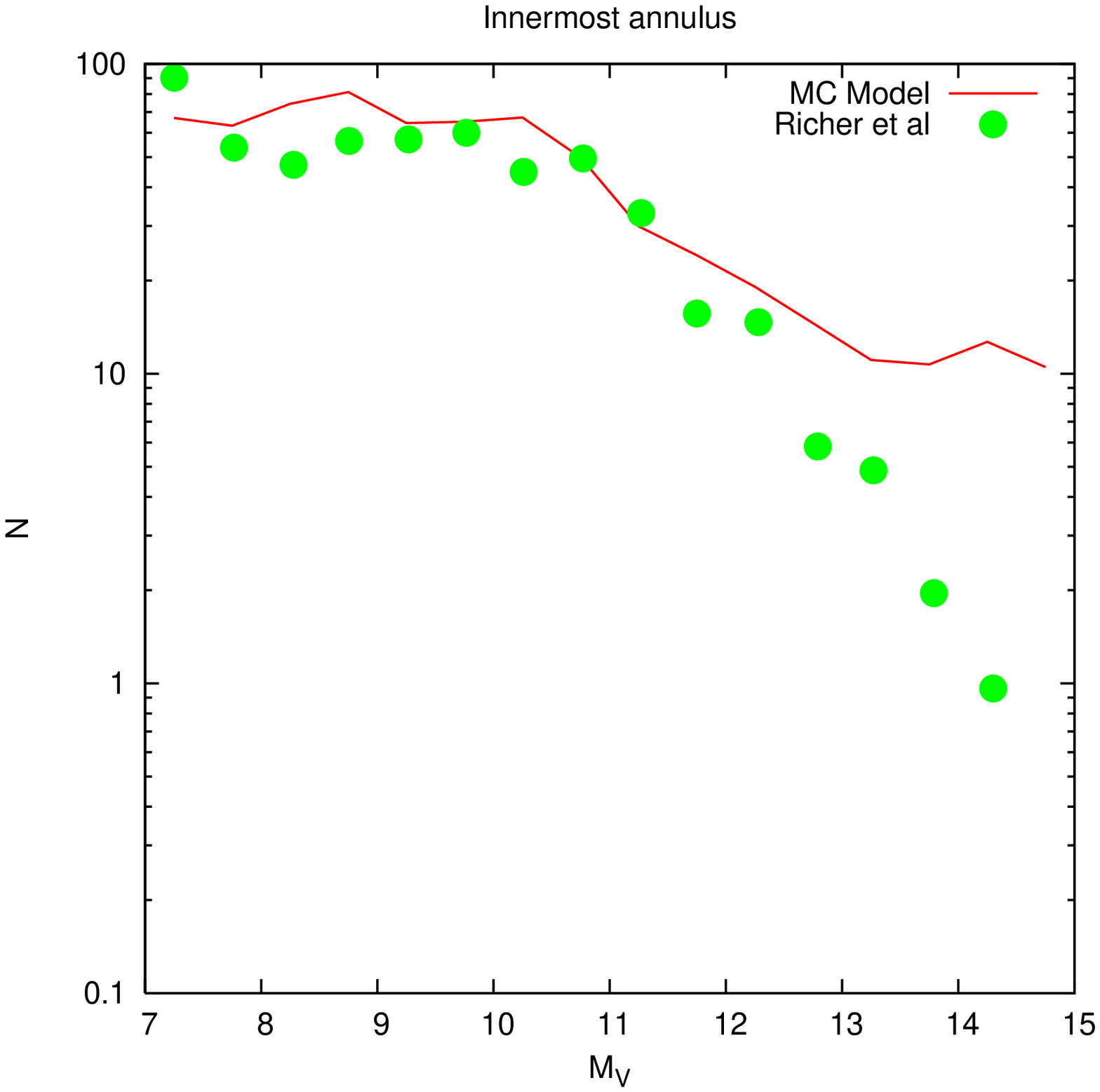}}
    \caption{Luminosity function  of our Monte Carlo model at the
    median radius of the innermost annulus in \citet{Ri2004}, compared with their data.}
    \end{minipage}
\hfill
    \begin{minipage}[]{0.45\textwidth}
{\includegraphics[height=10cm,angle=-90,width=8cm]{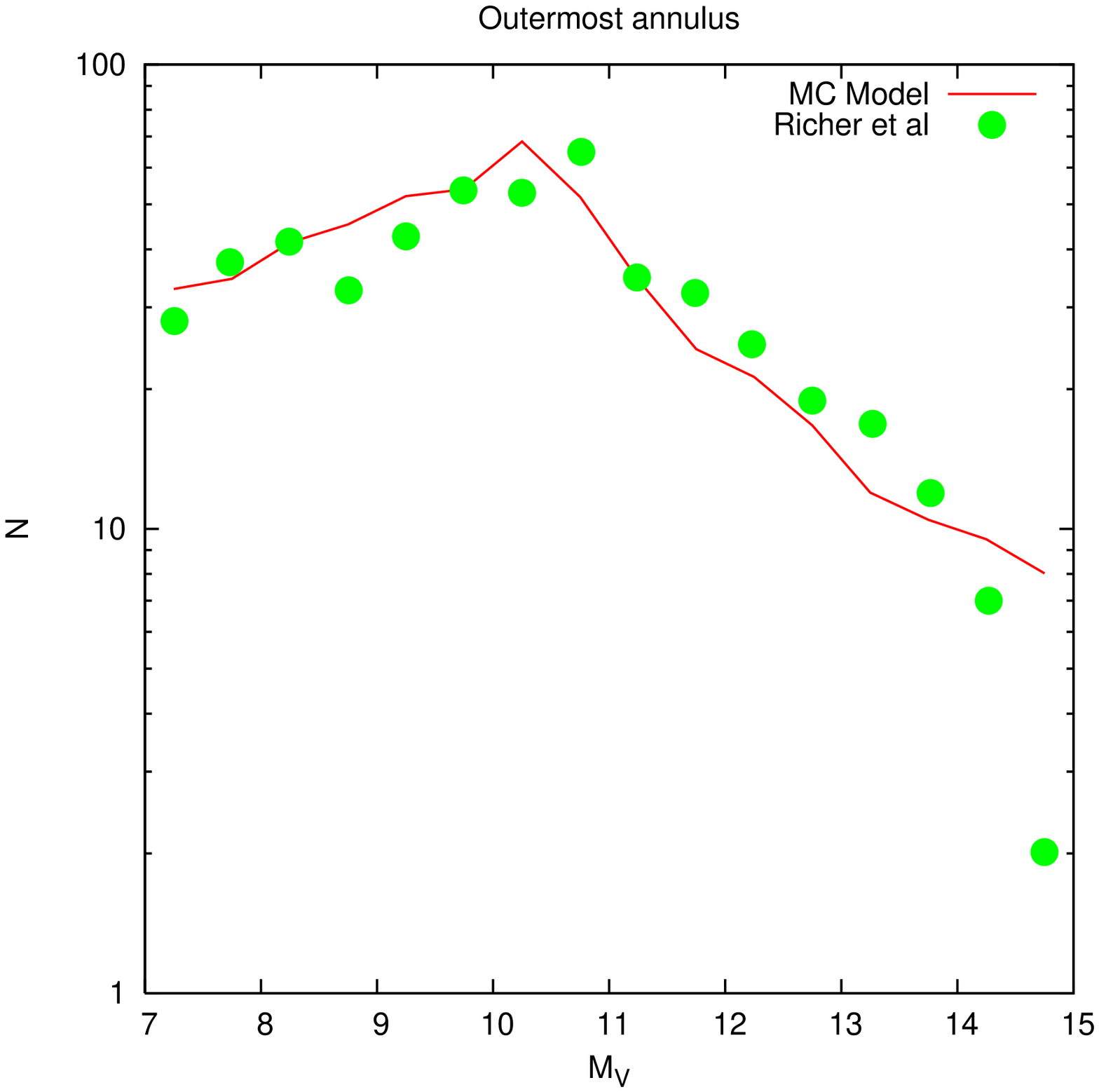}}
    \caption{Luminosity function  of our Monte Carlo model at the
    median radius of the outermost annulus in \citet{Ri2004}, compared with their data.}
    \end{minipage}
  \end{figure}

\begin{figure}
\begin{center}
\includegraphics[angle=-90,width=14cm]{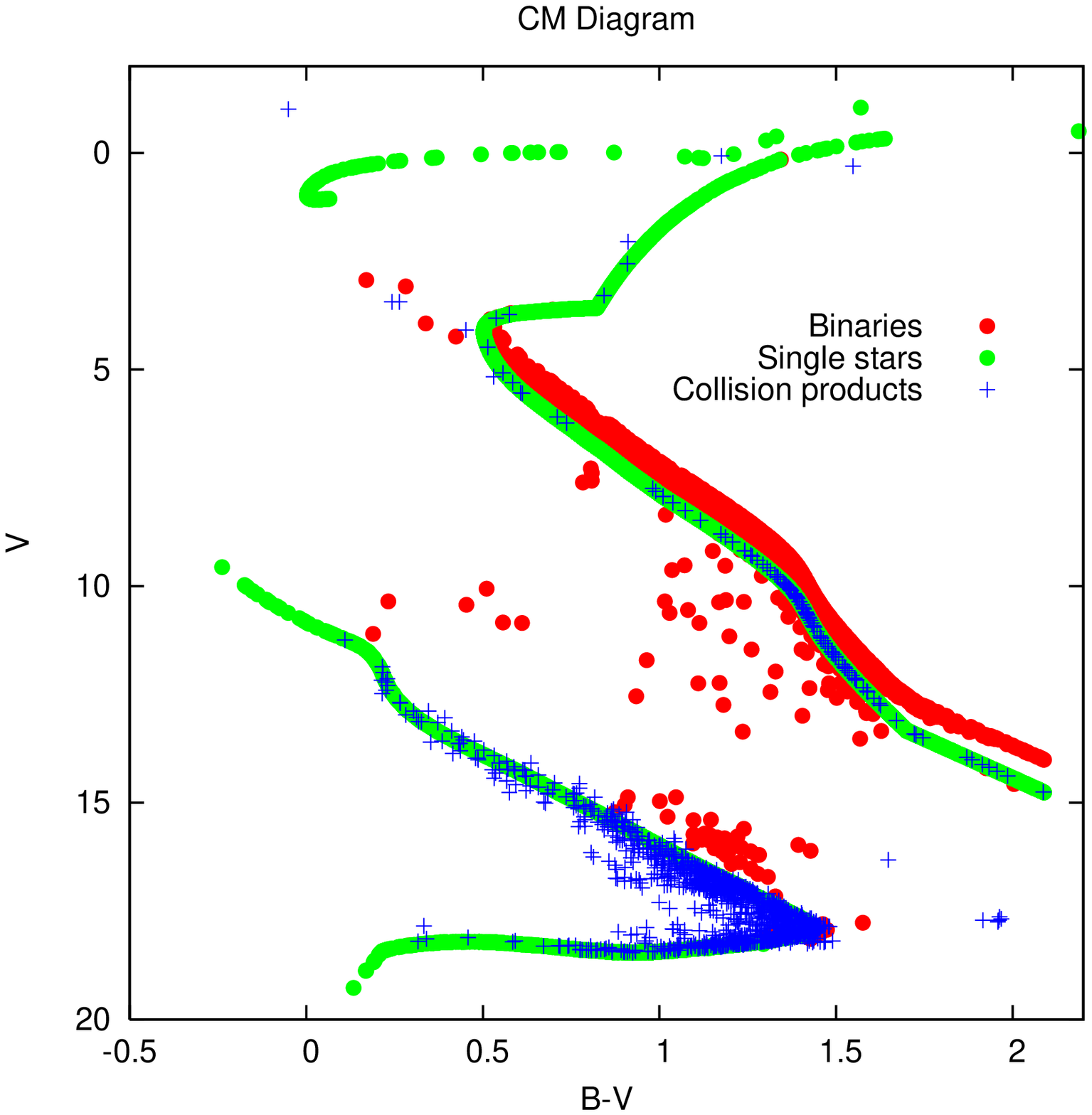}
     \caption{The colour-magnitude diagram at 12Gyr.  Green: single
 stars; red: binaries; blue pluses: collision or merger remnants.}
 \label{fig:colour-magnitude}
\end{center}
   \end{figure}

Figure \ref{fig:colour-magnitude} shows the colour-magnitude diagram of the model.
This is of interest, not so much for comparison with observations, but
for the presence of a number of interesting features.  The division of
the lower main sequence is simply an artifact of the way binary masses
were selected (a total mass above 0.2$\msun$ and a component mass
above 0.1$\msun$.)  Of particular interest are the high numbers of
merger remnants on the lower white dwarf sequence.  There are very few
blue stragglers.   Partly this is a result of the low binary
frequency, but it is also
important to note that some formation channels are unrepresented in
our models (in particular, collisions during triple or four-body
interactions, though if a binary emerges from an interaction with
appropriate parameters, it will be treated as merged.)  These numbers
also depend on the assumed initial distribution of semi-major axis,
which is not yet well constrained by observations in globular clusters.

\begin{figure}
    \begin{minipage}[]{0.45\textwidth}
{\includegraphics[height=10cm,angle=-90,width=8cm]{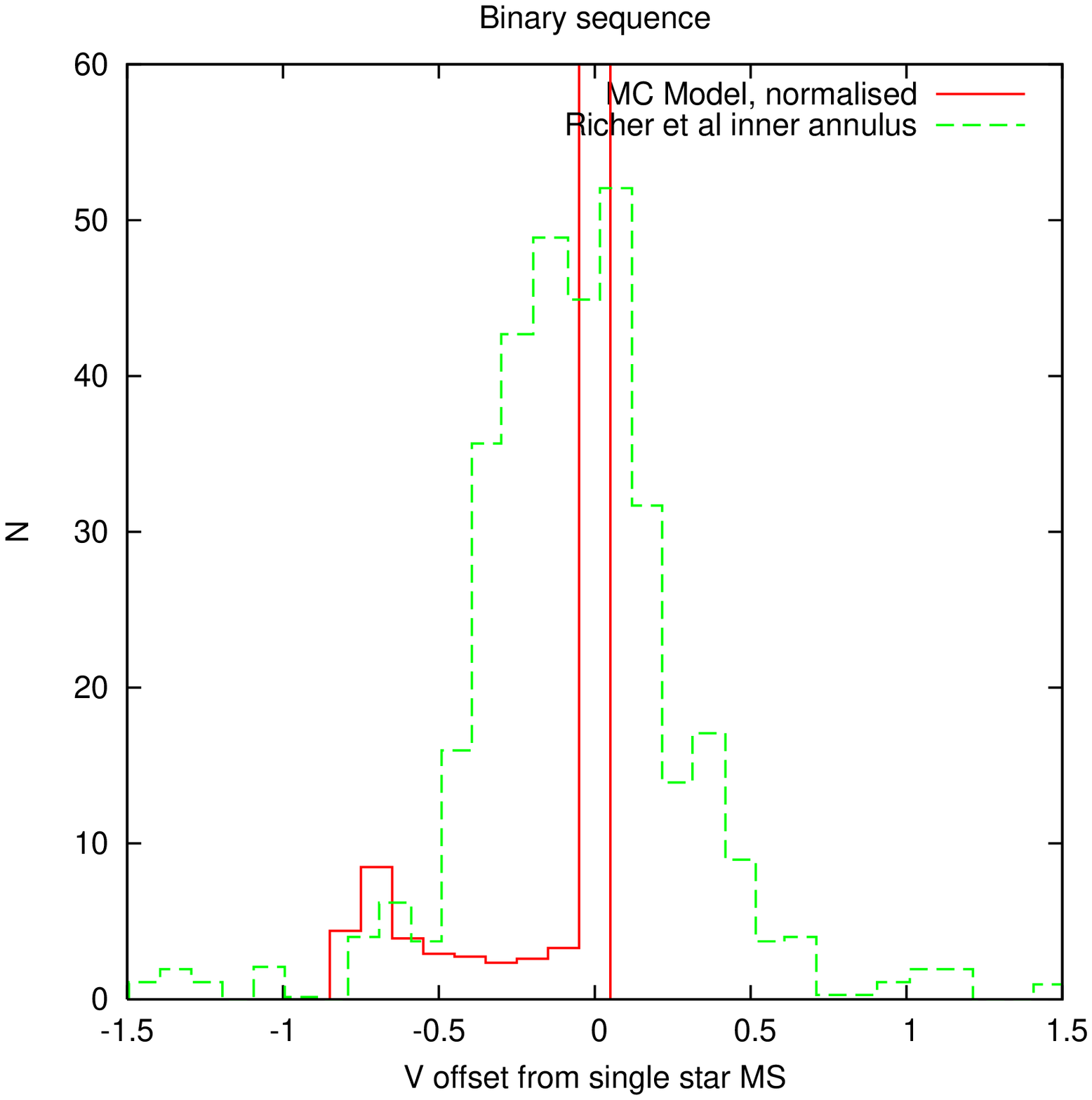}}
    \caption{Histogram of $V$-offset from the main sequence, compared
with the corresponding data from the innermost annulus studied by
Richer et al.  See text for details.}
\label{fig:photometrics}
    \end{minipage}
\hfill
\begin{minipage}[]{0.45\textwidth}
{\includegraphics[height=10cm,angle=-90,width=8cm]{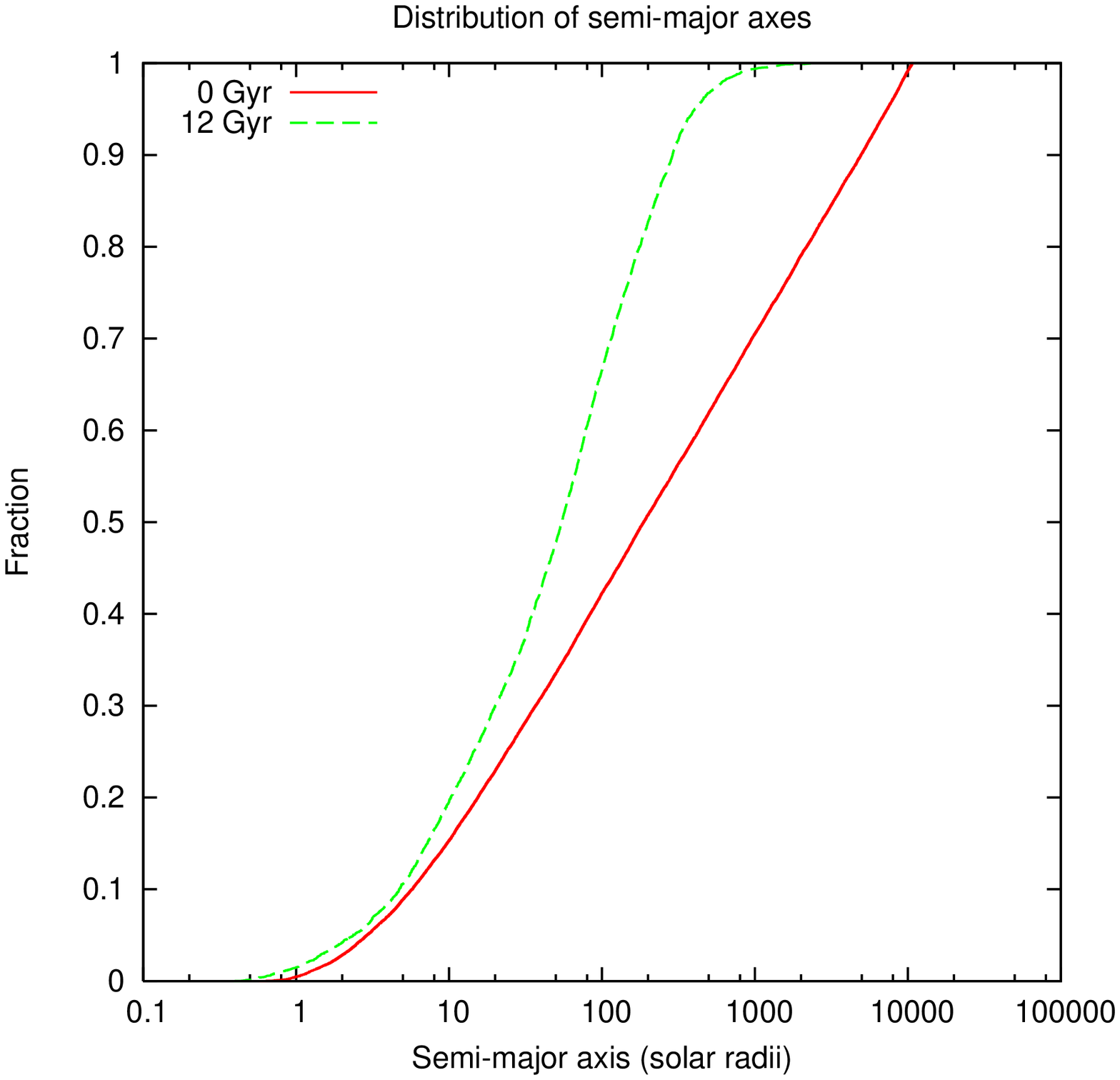}}
\caption{Distribution function of the semi-major axes of the
binaries at 0Gyr and 12Gyr.  Units: solar radii.}
\label{fig:sma}
\end{minipage}
  \end{figure}

Photometric binaries are visible in Fig.\ref{fig:colour-magnitude}, and these are
compared with observations in the inner field of \citet{Ri2004}
in Fig.\ref{fig:photometrics}.   In this figure, the model histogram has been normalised to the same
total number of stars as the observational one.  We made no attempt
to simulate photometric errors, but the bins around abscissa = -0.75
suggest that the binary fractions in the model and the observations
are comparable.  Fig.\ref{fig:sma} shows that binaries have evolved dynamically as well as through their internal
evolution.  In particular the softest pairs been
almost destroyed.

\begin{figure}
  \begin{minipage}[]{0.45\textwidth}
{\includegraphics[height=10cm,angle=-90,width=8cm]{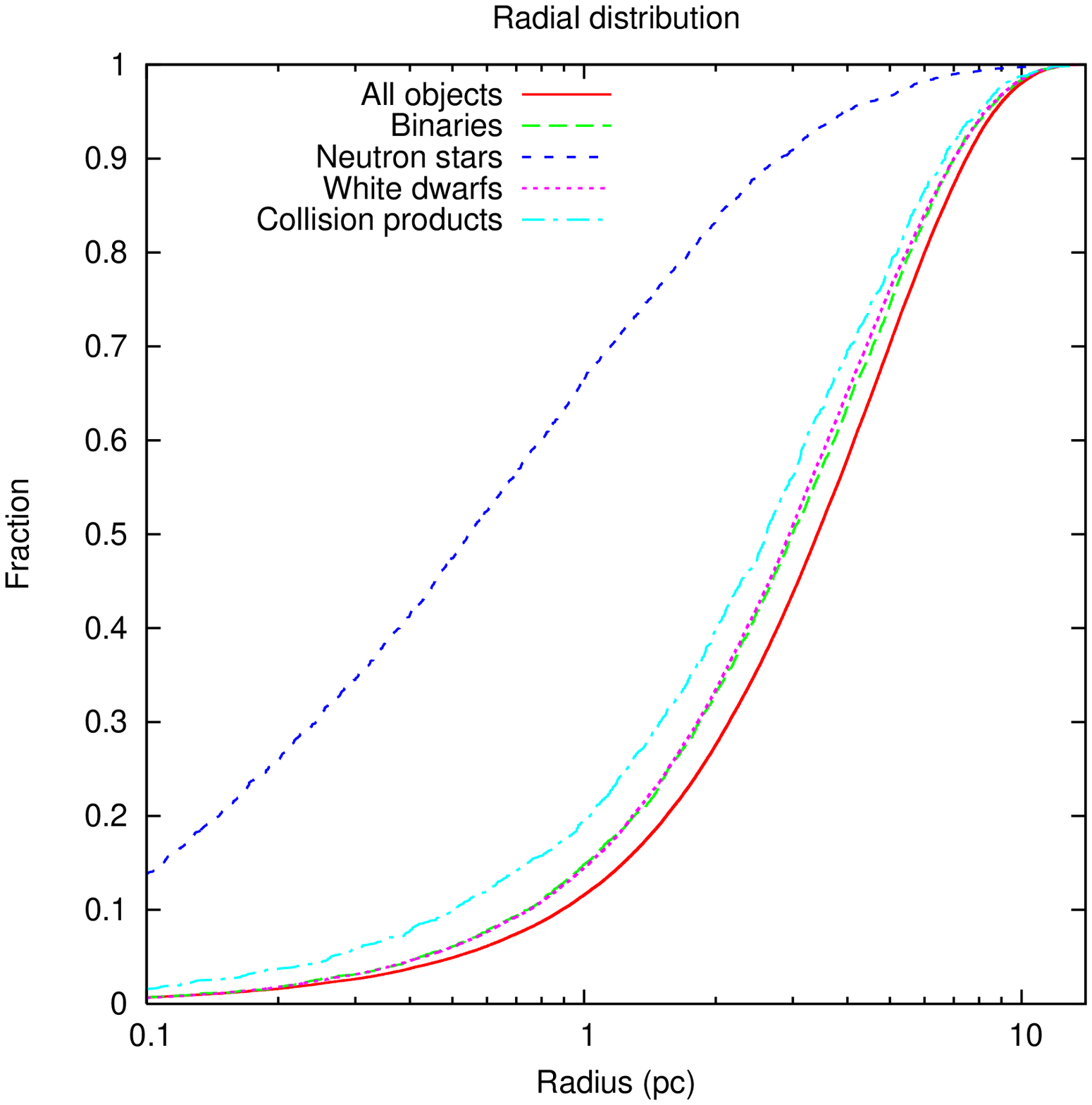}}
\caption{Radial distribution functions at 12Gyr.  Units: parsecs.  The
key identifies the class of object included.  The distributions of
white dwarfs and binaries are almost identical.}
\label{fig:segregation}
  \end{minipage}
\hfill
\begin{minipage}[]{0.45\textwidth}
{\includegraphics[height=10cm,angle=-90,width=8cm]{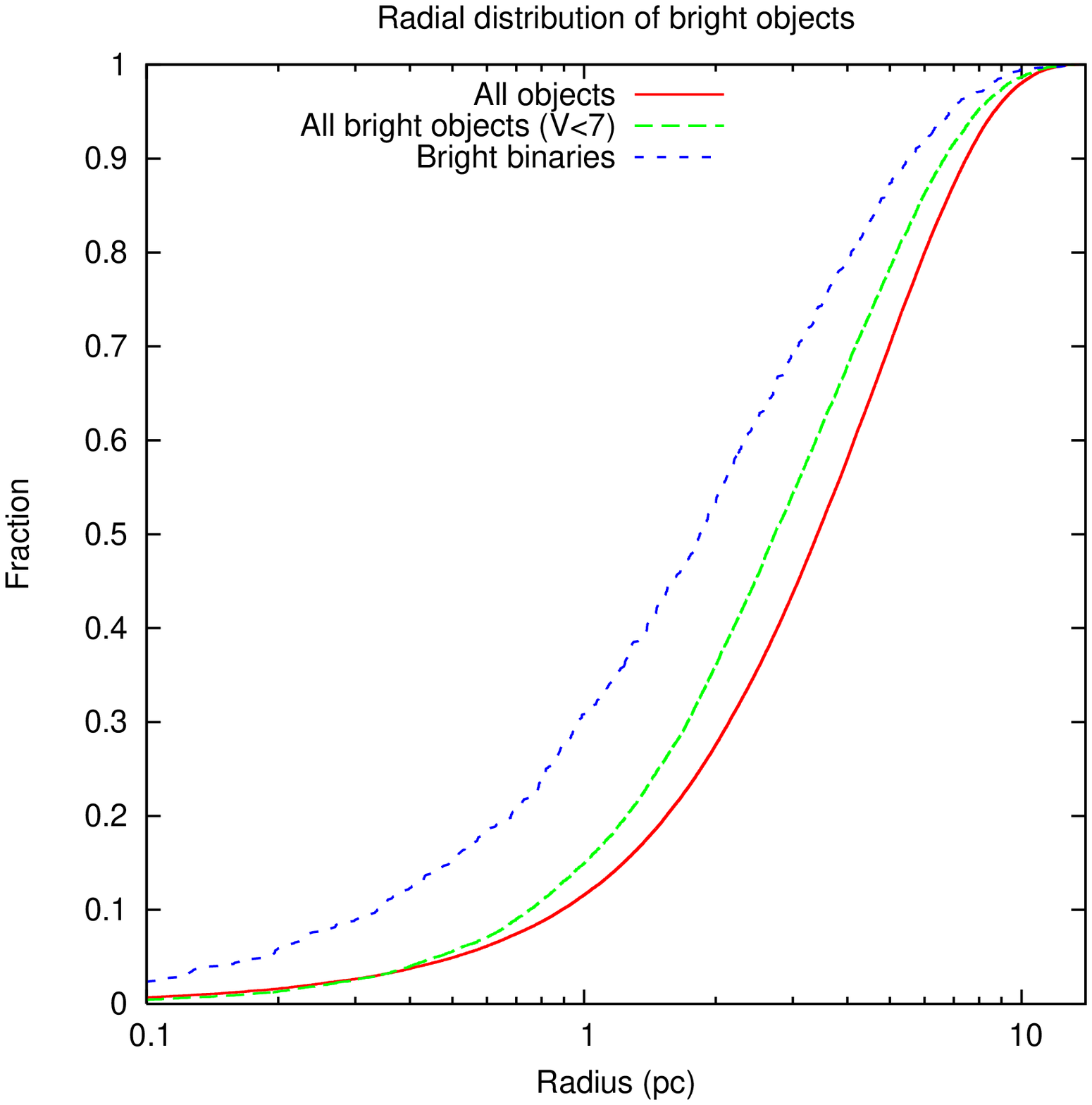}}
\caption{Radial distribution functions at 12Gyr, showing the extent of
segregation between bright single and bright binary stars
($M_V<7$). Also shown for comparison is the distribution for all
objects, as in Fig.\ref{fig:segregation}.  Units: parsecs.}
\label{fig:brights}
\end{minipage}
\end{figure}

By 12 Gyr the binaries exhibit segregation towards the centre of the
cluster, but perhaps in more subtle ways than might be expected
(Figs.\ref{fig:segregation},\ref{fig:brights}).  When {\sl all}
binaries are considered, there is little segregation relative to the
other objects in the system.  (Most binaries in our model are of low mass.)  But if one
restricts attention to bright binaries, which we here take to mean
those with $M_V < 7$ (i.e. brighter than about two magnitudes below
turnoff), the segregation is very noticeable (Fig.\ref{fig:brights}),
with a half-mass radius smaller by almost a factor of 2 than for bright
single stars.  Still, bright binaries are not nearly as
mass-segregated as neutron stars (Fig.\ref{fig:segregation}), which,
incidentally, receive no natal kicks in our model.

These data do not reveal one very interesting feature of our model,
which is that it exhibited core collapse at about 8Gyr.  Subsequently
its core radius is presumably sustained by binary burning.  Even
non-primordial binaries may be playing a role here.  To the
best of our knowledge it has not previously been suggested that M4, which is
classified as a ``King'' cluster by \citet{Tr1993}, is a post-collapse cluster.
This raises the long-dormant question of how it is possible for some
clusters to exhibit collapsed cores if they also come with significant
populations of primordial binaries.

\section{The search for initial conditions}

Each Monte Carlo model for this cluster takes a few days.  Therefore
the problem of finding appropriate initial conditions is a significant
one.  One needs a good starting guess, and then a rapid method for
iterative improvement.  Our techniques for dealing with these issues
are still primitive, but
have evolved in the course of this research.   At the iterative
stage we have employed scaled-up small models (with as few as $10^4$
objects sometimes), which are designed to relax at the same rate as a
full-scale model.  This requires a scaling of the length-scale, which
does violence to binary interactions, but nevertheless it has been
very useful.  In this section we focus on the issue of finding
starting values.

Consider first the problem of forwards evolution.  Useful formulae for
$M(t)$ are given by \citet{La2005}, which we have generalised to
other IMF's and metallicities.  To this we have added formulae for
$r_h(t)$, based on very simple notions of adiabatic expansion (in
response to mass-loss from stellar evolution) and tidal truncation.
For the evolution of the core we have fitted simple expressions to the
results on the time of core collapse given by \citet{BM2003}, extended by new $N$-body simulations to a wider range of
initial concentrations.  Very simple expressions for $r_c(t)$,
consistent with the initial and final values, can then be employed as
a first approximation.  Finally we drew from Baumgardt \& Makino a
relation between $M$ and $\alpha$ (the slope of the lower mass
function.)  Putting these relations together, we have constructed a
tool which we refer to as {\sl Quick Cluster Evolution}, following
S. Portegies Zwart.  To apply this to generate initial conditions for
our simulations, we can run QCE iteratively in reverse.

Further development of this tool should include the addition of binary
heating, which certainly influences the evolution of the half-mass
radius unless the primordial binary fraction is low enough, and more
concentrated initial conditions than the King models to which we have
restricted QCE so far.

\section{Discussion}

It is shown by Giersz \& Heggie (this volume) that Monte Carlo models can
provide similar results to $N$-body models, in the range where
comparison is possible, 
with similar physics (binaries, stellar evolution, etc.), except for a
number of restrictions:\begin{enumerate}
\item Use of a tidal cutoff, instead of the tidal field.  Though
  Sec.\ref{sec:mcm} summarises our current approach to this problem,  other treatments are possible, and worth
trying.
\item Use of a {\sl static} tide.  The effects of tidal shocks have
been studied by a number of authors \citep[e.g.][]{Ku1995}, and
it would be possible to add the effects as another process altering
the energies and angular momenta of the stars in the simulation.
\item Rotation: it has been shown \citep{Ki2004} that, to the extent
that rotating and non-rotating models can be compared, rotation
somewhat accelerates the rate of core collapse.  Rotation is hard to
implement in this Monte Carlo model, however.
\item Use of cross sections for triple/quad interactions: this
limitation could be overcome by direct integration of the
interactions, as is done by \citet{FR2007} in their version of the
Monte Carlo scheme.  It will be important to do so, as it would remove
the dependence on cross sections which are not well known for unequal
masses, and would also permit us to include the collisions between
stars which commonly occur in long-lived few-body interactions
\citep{HI1985}.
\item Neglect of triples:  these are also commonly produced in
binary-binary encounters \citep{Mi1984a}, and it is desirable to
include these as a third species (beyond single and binary stars).
Their observable effects may be small, but of course there is one
intriguing example in the very cluster we have focused on here \citep{Th1993}.
\end{enumerate}

Despite these limitations, not all of which are easily curable, Monte
Carlo models are feasible in reasonable time for globular clusters,
which are too large for direct $N$-body models.  They yield
predictions for mass segregation, luminosity functions, distributions
of binary parameters, anisotropy, and many other kinds of data, which
can hardly be obtained in any other way.  (The only comparable method
of which we are aware is the hybrid code of \citealt{GS2003}.)
Even when $N$-body
simulations eventually become possible, Monte Carlo models will remain as a quicker
way of exploring the main issues, just as King models have continued
to dominate the field of star cluster modelling even when more
advanced methods (e.g. Fokker-Planck models) have become available.

In addition to some of the possible improvements mentioned above, it
is our intention to extend the approach to a number of other objects,
including a ``collapsed-core'' cluster such as NGC6624.  We welcome
all suggestions for observational or theoretical comparisons, either
on M4 or on other objects.

\acknowledgements{J. Hurley's help has been invaluable,
especially in regard to stellar and binary evolution, and he went to
much trouble to assist us.  K. Meyer has been largely responsible for
developing QCE, following initial work with S. Portegies Zwart.  Her
work was funded by the Carnegie Trust for the Universities of
Scotland.  We thank H.B. Richer for comments on a previous draft.}


\begin{thebibliography}

\bibitem[Angeletti et al.(1980)]{An1980} Angeletti, L., 
Dolcetta, R., \& Giannone, P.\ 1980, Astrophys. Sp. Sci., 69, 45 

\bibitem[Baumgardt(2001)]{Ba2001} Baumgardt, H.\ 2001, \mnras, 
325, 1323

\bibitem[Baumgardt et al.(2003)]{Ba2003} Baumgardt, H., Hut, 
P., Makino, J., McMillan, S., \& Portegies Zwart, S.\ 2003, \apjl, 582, L21 

\bibitem[Baumgardt \& Makino(2003)]{BM2003} Baumgardt, H., \& 
Makino, J.\ 2003, \mnras, 340, 227 

\bibitem[Dinescu et al.(1999)]{Di1999} Dinescu, D.~I., Girard, 
T.~M., \& van Altena, W.~F.\ 1999, \aj, 117, 1792

\bibitem[Drukier(1993)]{Dr1993} Drukier, G.~A.\ 1993, \mnras, 
265, 773 

\bibitem[Drukier(1995)]{Dr1995} Drukier, G.~A.\ 1995, \apjs, 
100, 347 

  \bibitem[Drukier et al.(1988)]{Dr1988} Drukier, G.~A., 
Fahlman, G.~G., Richer, H.~B., \& Vandenberg, D.~A.\ 1988, \aj, 95, 1415 

\bibitem[Dubath et al.(1990)]{Du1990} Dubath, P., Meylan, G., 
Mayor, M., \& Magain, P.\ 1990, \aap, 239, 142

\bibitem[Dull et al.(1997)]{Du1997} Dull, J.~D., Cohn, H.~N., 
Lugger, P.~M., Murphy, B.~W., Seitzer, P.~O., Callanan, P.~J., Rutten, 
R.~G.~M., \& Charles, P.~A.\ 1997, \apj, 481, 267 

\bibitem[Fleck et al.(2006)]{Fl2006} Fleck, J.-J., Boily, 
C.~M., Lan{\c c}on, A., \& Deiters, S.\ 2006, \mnras, 369, 1392 

\bibitem[Fregeau \& Rasio(2007)]{FR2007} Fregeau, J.~M., \& 
Rasio, F.~A.\ 2007, \apj, 658, 1047

\bibitem[Gebhardt \& Fischer(1995)]{GF1995} Gebhardt, K., \& 
Fischer, P.\ 1995, \aj, 109, 209 

\bibitem[Giersz(1998)]{Gi1998} Giersz, M.\ 1998, \mnras, 298, 
1239 

\bibitem[Giersz(2001)]{Gi2001} Giersz, M.\ 2001, \mnras, 324, 
218 

\bibitem[Giersz(2006)]{Gi2006} Giersz, M.\ 2006, \mnras, 371, 
484 

\bibitem[Giersz \& Heggie(2003)]{GH2003} Giersz, M., \& 
Heggie, D.~C.\ 2003, \mnras, 339, 486 

\bibitem[Giersz \& Spurzem(2003)]{GS2003} Giersz, M., \& 
Spurzem, R.\ 2003, \mnras, 343, 781

\bibitem[Grabhorn(1993)]{Gr1993} Grabhorn, R.~P.\ 1993, 
Ph.D.~Thesis,  Indiana University

\bibitem[Grabhorn et al.(1992)]{Gr1992} Grabhorn, R.~P., Cohn, 
H.~N., Lugger, P.~M., \& Murphy, B.~W.\ 1992, \apj, 392, 86 

\bibitem[Gunn \& Griffin(1979)]{GG1979} Gunn, J.~E., \& 
Griffin, R.~F.\ 1979, \aj, 84, 752 

\bibitem[Hansen et al.(2002)]{Ha2002} Hansen, B.~M.~S., et 
al.\ 2002, \apjl, 574, L155 

\bibitem[Hansen et al.(2004)]{Ha2004} Hansen, B.~M.~S., et 
al.\ 2004, \apjs, 155, 551

\bibitem[Harris(1996)]{Ha1996} Harris, W.~E.\ 1996, \aj, 112, 
1487

\bibitem[Heggie et al.(2006)]{He2006} Heggie, D.~C., Portegies 
Zwart, S., \& Hurley, J.~R.\ 2006, New Astronomy, 12, 20

\bibitem[Hurley et al.(2000)]{Hu2000} Hurley, J.~R., Pols, 
O.~R., \& Tout, C.~A.\ 2000, \mnras, 315, 543

\bibitem[Hurley et al.(2002)]{Hu2002} Hurley, J.~R., Tout, 
C.~A., \& Pols, O.~R.\ 2002, \mnras, 329, 897 




\bibitem[Hurley et al.(2005)]{Hu2005} Hurley, J.~R., Pols, 
O.~R., Aarseth, S.~J., \& Tout, C.~A.\ 2005, \mnras, 363, 293

\bibitem[Hurley(2007)]{Hu2007} Hurley, J.~R.\ 2007, \mnras, 
379, 93 

\bibitem[Hut \& Inagaki(1985)]{HI1985} Hut, P., \& Inagaki, 
S.\ 1985, \apj, 298, 502

\bibitem[Illingworth(1976)]{Il1976} Illingworth, G.\ 1976, 
\apj, 204, 73 

\bibitem[Kim et al.(2004)]{Ki2004} Kim, E., Lee, H.~M., \& 
Spurzem, R.\ 2004, \mnras, 351, 220

\bibitem[King(1966)]{Ki1966} King, I.~R.\ 1966, \aj, 71, 64 

\bibitem[Kroupa(1995)]{Kr1995} Kroupa, P.\ 1995, \mnras, 277, 
1507 

\bibitem[Kroupa(2007)]{Kr2007} Kroupa, P.\ 2007, ArXiv 
Astrophysics e-prints, arXiv:astro-ph/0703124 

\bibitem[Kroupa et al.(1991)]{Kr1991} Kroupa, P., Gilmore, G., 
\& Tout, C.~A.\ 1991, \mnras, 251, 293

\bibitem[Kroupa et al.(2001)]{Kr2001} Kroupa, P., Aarseth, S., 
\& Hurley, J.\ 2001, \mnras, 321, 699 

\bibitem[Kundic \& Ostriker(1995)]{Ku1995} Kundic, T., \& 
Ostriker, J.~P.\ 1995, \apj, 438, 702

\bibitem[Lamers et al.(2005)]{La2005} Lamers, H.~J.~G.~L.~M., 
Gieles, M., Bastian, N., Baumgardt, H., Kharchenko, N.~V., \& Portegies 
Zwart, S.\ 2005, \aap, 441, 117

\bibitem[Leonard et al.(1992)]{Le1992} Leonard, P.~J.~T., 
Richer, H.~B., \& Fahlman, G.~G.\ 1992, \aj, 104, 2104

\bibitem[Michie \& Bodenheimer(1963)]{MB1963} Michie, R.~W., 
\& Bodenheimer, P.~H.\ 1963, \mnras, 126, 269 

\bibitem[Mikkola(1983)]{Mi1983} Mikkola, S.\ 1983, \mnras, 
205, 733

\bibitem[Mikkola(1984a)]{Mi1984a} Mikkola, S.\ 1984a, \mnras, 
207, 115 

\bibitem[Mikkola(1984b)]{Mi1984b} Mikkola, S.\ 1984b, \mnras, 
208, 75 

\bibitem[Peterson \& King(1975)]{PK1975} Peterson, C.~J., \& 
King, I.~R.\ 1975, \aj, 80, 427 

\bibitem[Peterson et al.(1995)]{Pe1995} Peterson, R.~C., Rees, 
R.~F., \& Cudworth, K.~M.\ 1995, \apj, 443, 124

\bibitem[Phinney(1993)]{Ph1993} Phinney, E.~S.\ 1993, in Djorgovski,
  S.G., Meylan G., eds, 
Structure and Dynamics of Globular Clusters, ASPCS 50, 141 

\bibitem[Plummer(1911)]{Pl1911} Plummer, H.~C.\ 1911, \mnras, 
71, 460 

\bibitem[Portegies Zwart \& Verbunt(1996)]{PZ1996} Portegies 
Zwart, S.~F., \& Verbunt, F.\ 1996, \aap, 309, 179 


\bibitem[Pryor et al.(1986)]{Pr1986} Pryor, C., Hartwick, 
F.~D.~A., McClure, R.~D., Fletcher, J.~M., \& Kormendy, J.\ 1986, \aj, 91, 
546 

\bibitem[Richer et al.(2004)]{Ri2004} Richer, H.~B., et al.\ 
2004, \aj, 127, 2771 

\bibitem[Spitzer(1987)]{Sp1987} Spitzer, L.\ 1987, Princeton, 
NJ, Princeton University Press, 1987

\bibitem[Thorsett et al.(1993)]{Th1993} Thorsett, S.~E., 
Arzoumanian, Z., \& Taylor, J.~H.\ 1993, \apjl, 412, L33

\bibitem[Trager et al.(1993)]{Tr1993} Trager, S.~C., 
Djorgovski, S., \& King, I.~R.\ 1993, in Djorgovski,
  S.G., Meylan G., eds, Structure and Dynamics of Globular 
Clusters, ASPCS 50, 347

\bibitem[Trager et al.(1995)]{Tr1995} Trager, S.~C., King, 
I.~R., \& Djorgovski, S.\ 1995, \aj, 109, 218

\bibitem[van de Ven et al.(2006)]{Ve2006} van de Ven, G., van 
den Bosch, R.~C.~E., Verolme, E.~K., \& de Zeeuw, P.~T.\ 2006, \aap, 445, 
513 

\bibitem[van den Bosch et al.(2006)]{Bo2006} van den Bosch, 
R., de Zeeuw, T., Gebhardt, K., Noyola, E., \& van de Ven, G.\ 2006, \apj, 
641, 852 

\bibitem[Von Zeipel(1908)]{Ze1908} Von Zeipel, H.\ 1908, 
Annales de l'Observatoire de Paris, 25, 1 

\end{thebibliography}
\end{document}